# Influence of ground motion duration on the structural response at multiple seismic intensity levels

Mojtaba Harati[1], Mohammadreza Mashayekhi[2], Morteza Ashoori Barmchi[2] and Homayoon E. Estekanchi[2]

[1]Department of Civil Engineering, University of Science and Culture, Rasht, Iran
[2]Department of Civil Engineering, Sharif University of Technology, Tehran, Iran

**Abstract**

This paper aims to investigate the effects of motion duration on the structural seismic demands, seeking potential correlations between motion durations and structural responses at several seismic intensity levels. Three seismic levels with 100years, 475years, and 2475years earthquake return periods (RPs) are first considered for correlation computations. Spectrally matched ground motions are employed to isolate the contribution of duration from the effects of ground motion amplitudes and response spectral shape. Four single degree of freedom systems derived from four real reinforced concrete structures are studied, where both degrading and non-degrading equivalent SDOF systems are included for structural modeling. Results show a low positive correlation between motion duration and structural displacement demand but this correlation increases with an increase in earthquake RP. It is also investigated whether or not this small positive correlation has an impact on the incremental dynamic analysis curves. The spectrally matched ground motions are divided into two distinct groups in this case: short- and long-duration ground motions. The comparison of incremental dynamic analysis of these two groups at the collapse limit reveals that long-duration ground motions can cause up to a 20 percent decrease in the spectral acceleration demand of considered structural systems.

---

Corresponding: Mohammadreza Mashayekhi, Research Associate, Sharif University of Technology, Tehran, Iran. Email: mmashayekhi67@gmail.com

## 1 Introduction

Post-earthquake field reports show that damages in the structural members of the observed building type systems can be pertinent to the duration of the induced earthquakes as well as high nonlinear cycles endured by the elements of the damaged structural systems before the failure. It is also established by a growing body of the research in this area that earthquake durations may have a meaningful effect on the structural performance of built infrastructures. Numerous researchers have worked on the seismic response of different structures regarding the influence of motion duration. Their studies revealed that seismic responses of the structures under earthquake loadings with deteriorative behaviors, including RC frames [1–5], concrete dams [6–8], SDOF systems with pinching-degrading behavior [9] and masonry buildings [10], are directly influenced by the duration of ground motions. Timber structures are also influenced by ground motion duration, as presented by Pan et al. [11,12]. This means that structures with degrading behaviors are more vulnerable to motion duration, so more structural and non-structural damages would be expected to occur at places where constructions may be subjected to long-duration earthquakes [10,12]. As a result, accumulated damage indices which are partially or completely composed of the hysteretic cyclic energy of the earthquakes such as Pak-Ang damage index [14] as well as extreme damage indices such as peak floor drifts are shown to have positive correlations with the motion durations [5,12,13].

Although it is shown that there is a positive correlation between motion duration and different damage induces, current seismic codes generally offer a record selection procedure through which ground motions are mainly selected in such a way that their response spectrum is adequately compatible with a predefined target response spectrum [17]. In this case, some rules are prescribed by the codes to ensure the aforementioned response spectrum compatibility. Moreover, the minimum duration length of the earthquake is not explicitly dictated or even recommended by many seismic codes around the world—such as ASCE07 [18] and current US rehabilitation provisions (e.g. ASCE/SEI 41-17 [19] and FEMA-356 [20]). However, there are some seismic provisions that put a limit on the required minimum length of motion duration for use in the response time history analysis. For example, Chinese Code for Seismic Design of Buildings [21] necessitates a specified bracketed motion duration which is equal to or more than 5–10 times of the fundamental period of the structure. Also, a similar regulation can be found in Iranian National

Building Code (INBC) by which structural engineers are forced to select earthquake records with strong motion duration at least equal to 10 seconds or more than 3 times of the fundamental period of the structure. It is important to note that the minimum strong motion duration recommended by INBC can be of any definitions for the duration of the earthquakes.

In contrast to the above-mentioned studies that mainly focused on the correlation of motion duration and damage indices considering a specific hazard level—for example, the design basis earthquake (DBE) level or the maximum considered earthquake (MCE) level, this paper attempts to show the effects of motion duration on the structural seismic demands at different seismic hazard levels up to a point where a complete collapse of the structures occurs. Therefore, our study includes different hazard levels which are meaningful when structural seismic performance is the case. For this purpose, statistical correlation computations are considered to include the potential effects of considering different seismic hazard levels in such investigations.

## 2 Research methodology

In this paper, first the correlation of motion duration with structural seismic demands—for building type systems with and without degrading behavior—is considered using a devised research framework which is based on spectrally matched ground motions that are uniformly adjusted to be at the same seismic levels. The structural seismic demands required for correlation assessments are calculated by an Incremental Dynamic Analysis (IDA) [22] as well as a nonlinear response time history procedures. The record selection procedure essential for these analyses is prepared based on the regulations posed by INBC. In this case, the selected records are matched to a target spectrum to diminish their variability related to frequency content as well as spectral amplitudes. In order to diminish uncertainties and variability associated with the duration length of the ground motions and include its effect on the response analysis (or on the median IDA curves), earthquake ground motions are divided into two distinct groups—namely the short- and long-duration sets. Details about spectral matching procedure and policy regarding selected duration definition and the division of the records into two groups are given in the next sections.

These numerical dynamic analyses are performed on the equivalent SDOFs which are subjected to scaled matched ground motions. In this case, all adjusted ground motions, both from short- and long-duration sets of motions, are scaled to the desired seismic levels. Moreover, earthquake RPs serve as a means to change the levels of intensity measures (IM) in the response analysis. Statistical correlation procedures are performed at three distinct levels of seismic excitations. Each of these seismic levels corresponds to an earthquake RP. For example, earthquake RPs of 100, 475, and 2475 years represent the service level earthquake (SLE), design basis earthquake (DBE) and the maximum considered earthquake (MCE), respectively. RPs and scaling factors for the ground motions are obtained from hazard curves of the design spectrum of INBC at different vibration periods and seismic zones considering four different soil classes defined in the code [23], as indicated in Figure 1. In this figure, sample hazard curves have been obtained for a site located in a high seismic zone and the spectral acceleration at a vibration period of 0.5 seconds. It is of the essence to mention that different soil classes have been utilized for the derivation of these hazard curves. These soil types that are categorized based on the time-averaged shear wave velocity over a sub-surface depth 30 meters—the so-called Vs30 parameter—can include: hard rocks (Type 1 & 2), very dense soil and soft rock (Type 3), stiff soil (Type 4) and soft soil (Type 5). A region within a high seismic zone and the soil class 3—which is quite compatible with the soil category of the considered site in this study—are chosen for the calculation of RPs. It is essential to add that equivalent SDOFs are modeled utilizing standard pushover curves of multi-degree RC buildings, as suggested by Vamvatsikos and Cornell [24]. Characteristics of these SDOFs are also mentioned in the following section of the paper.

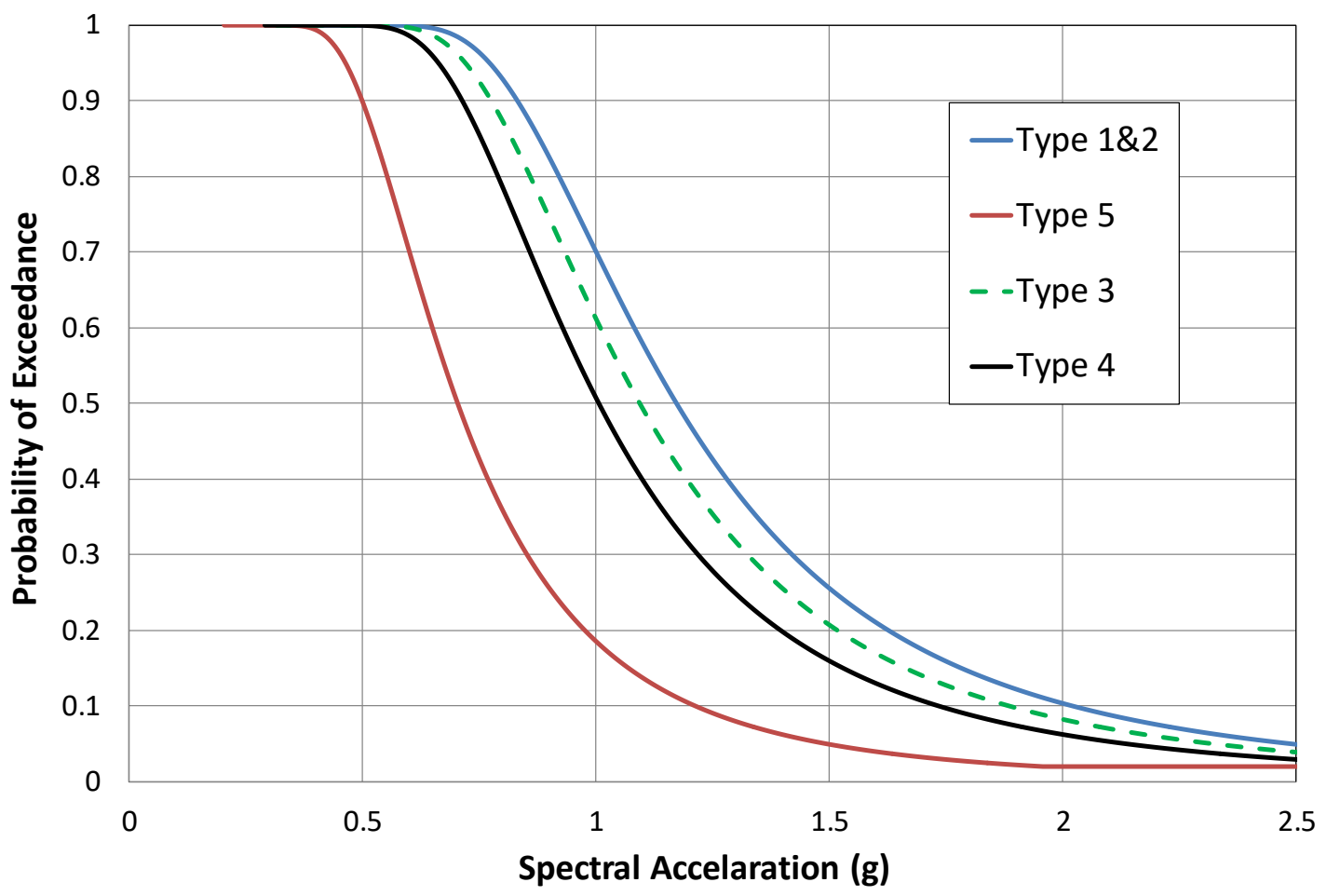

Figure 1. Sample hazard curves of INBC design spectrum for different soil types at a high seismic zone and a natural period of vibration equal to 0.5 seconds.

While two measurements of interest—the seismic demands at specific seismic level and duration of the selected motions—are independently generated and found in each computer simulation, it is possible to estimate the related correlation coefficient using the Pearson product-moment correlation estimator:

$$\rho_{x,y} = \frac{\sum_n[(x_i - \bar{x})(y_i - \bar{y})]}{\sqrt{\sum_n[(x_i - \bar{x})^2]\sum_n[(y_i - \bar{y})^2]}} \tag{1}$$

where $x_i$ and $y_i$ are the components of vector *X* and *Y*, which are related to the two measurements of interest respectively; x and y are the means of vector *X* and *Y*, and $\sum_n[\,]$ represents summation over the number of items pertinent to the duration utilized motions or over components of the vector related to the seismic demands obtained for the whole number of applied ground motions all of which are at a specific seismic hazard levels.

## 3 Definitions of motion duration

There are more than 30 definitions for motion duration—or duration of strong ground motion—in the literature as reported in [25], but some of them are more commonly accepted and used by the earthquake engineering community. Among the defined available definitions in the literature, bracketed duration, uniform duration as well as the significant duration are more repeatedly used in the field of earthquake engineering. The bracketed duration of motion delivers the total time left between the first and last acceleration excursions which are greater than a specific predefined threshold. The definition pertinent to the uniform duration is all related to the sum of the elapsed time intervals considering the same aforementioned threshold level set on the acceleration [26]. But the definition related to significant duration is somehow different from the bracketed and uniform duration. This definition of the motion duration takes use of a well-known integration-based accumulative intensity measure, the so-called Arias Intensity (AI). Significant duration is denoted by $D_{x\text{-}y}$ hereafter, which is defined as the time interval during which the normalized AI moves from a minimum ($x_{\%}$) to a maximum ($y_{\%}$) threshold. And so, the $D_{5\text{-}95}$ means the time interval as buildup accumulation energy of the earthquake goes up from 5 to 95 percent. It is of the essence to add that some studies show that the CAV can be also considered as an alternative for the AI to assess the effect of the motion duration on structural responses (e.g., EPRI [27], Cabañas et al. [28]). This is due to the fact that both of these intensity measures, the CAV and AI, are capable of capturing and showing the cumulative energy of the ground motions.

Both of the CAV and AI are defined as the time integral of a form of acceleration function profile as can be seen in Equation (2) and (3), where the $|a(t)|$ is the absolute value of the acceleration function of the ground motion at time $t$, $[a(t)]$. Also, $t_{max}$ and AI is the total duration of ground motion and the total AI calculated for the entire duration of the ground shakings.

$$AI = \frac{\pi}{2g} \int_0^{t_{max}} [a(t)]^2 \, dt \qquad (2)$$

$$CAV = \int_0^{t_{max}} |a(t)| \, dt \qquad (3)$$

While there are many definitions for the motion duration as indicated before, the definition for the significant duration is selected as a duration-related parameter in this paper because it is a continuous time interval as far as the characteristics of ground acceleration are concerned; therefore, this definition for duration of an earthquake is more convenient for the time history analysis. The process pertinent to the calculation of a form of significant duration, the $D_{5\text{-}95}$ parameter, for the Loma-Prieta earthquake of 1989 is depicted in Figure 2. According to the figure and as mentioned earlier, the significant duration ($D_{5\text{-}95}$) is the time interval during which the buildup energy of the normalized AI moves from a minimum (5%) to a maximum (95%) threshold. The times associated with the mentioned minimum and maximum thresholds are defined by $t_x$ (here 8sec) and $t_y$ (here 17sec), respectively.

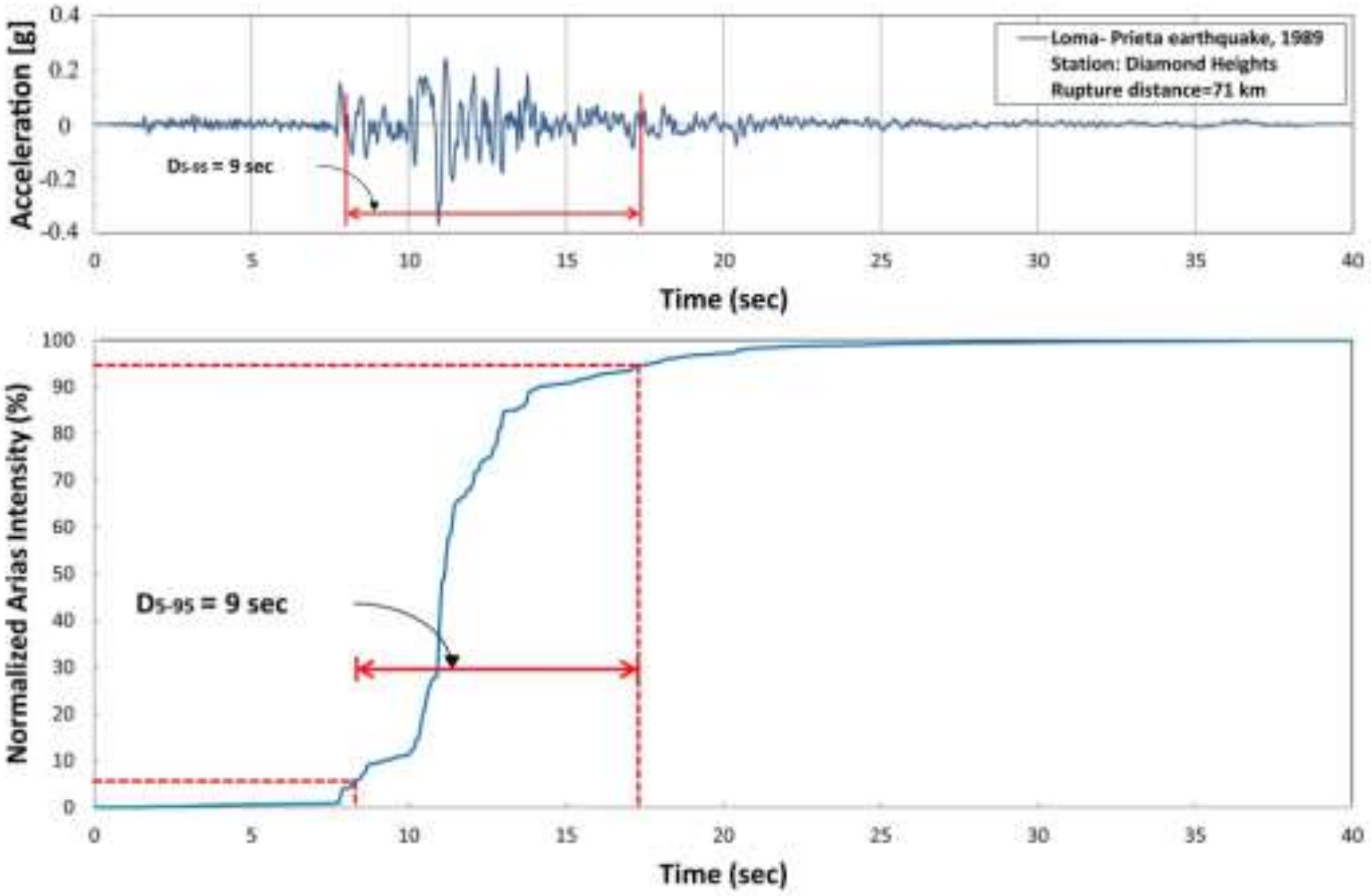

Figure 2. The procedure required to compute the D5-95 parameter of a recorded ground motion

## 4 Selection of earthquake records

### 4.1 Characteristics of the selected motions

First, a bin of ground motions (200 earthquake records)—which were selected by Heo et al. [29]—is considered as a source of the record selection procedure in this study. Next, a number of these 200 ground motions is first nominated and then divided into two different subsets: short- and long-duration sets. This record selection and division procedure are based on a devised mechanism which is described in section 4.3. Heo et al. [29] employed the aforementioned 200 earthquake motions in the dynamic analysis which aimed to compare amplitude scaled and spectrum-matched ground motions for seismic performance assessment. In this case, a selection of unscaled ground motions that can potentially drive considered structures into the nonlinear range was of interest. Therefore, a subset of 200 ground motions from Pacific Earthquake Engineering Research (PEER) Next Generation Attenuations (NGA), whose PGA exceeded 0.2g was used in their computer simulation. The magnitude-distance distribution of the selected earthquake records is extracted from Heo [30] and plotted in Figure 3. It contains information related to the two components of 100 earthquakes. Soil type C and D denotes to the soft and deep stiff soils, respectively. For further detailed explanation about the data reflected in Figure 3, please refer to Heo [30].

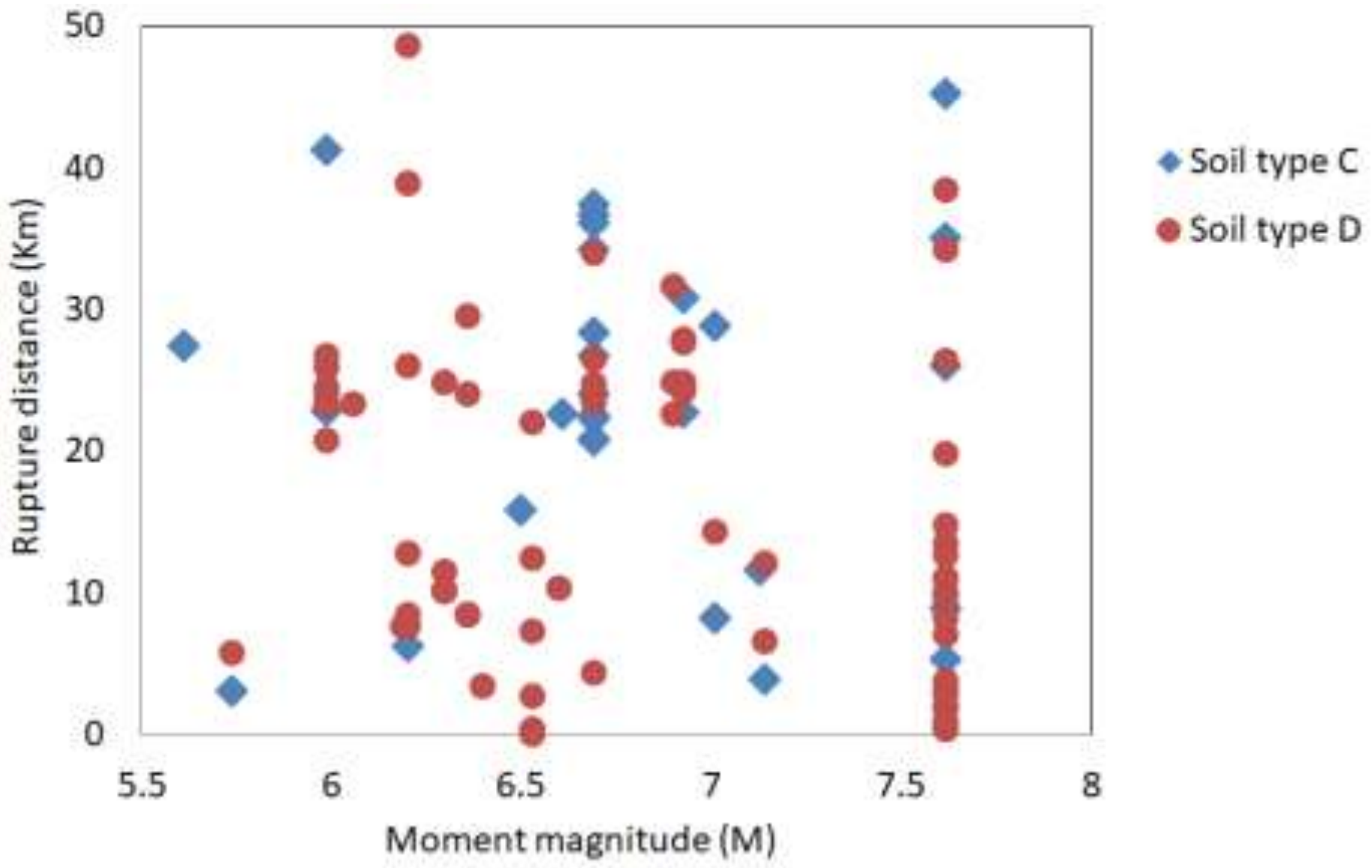

Figure 3. Distribution of magnitude-distance of the dataset [31]

Acceleration spectra of these ground motions are shown in Figure 4. As can be seen in this figure, dispersion of acceleration spectra of these ground motions is very noticeable which can also cause a considerable dispersion in the structural responses. This variability can be attributed to the amplitude and frequency content of the ground motions.

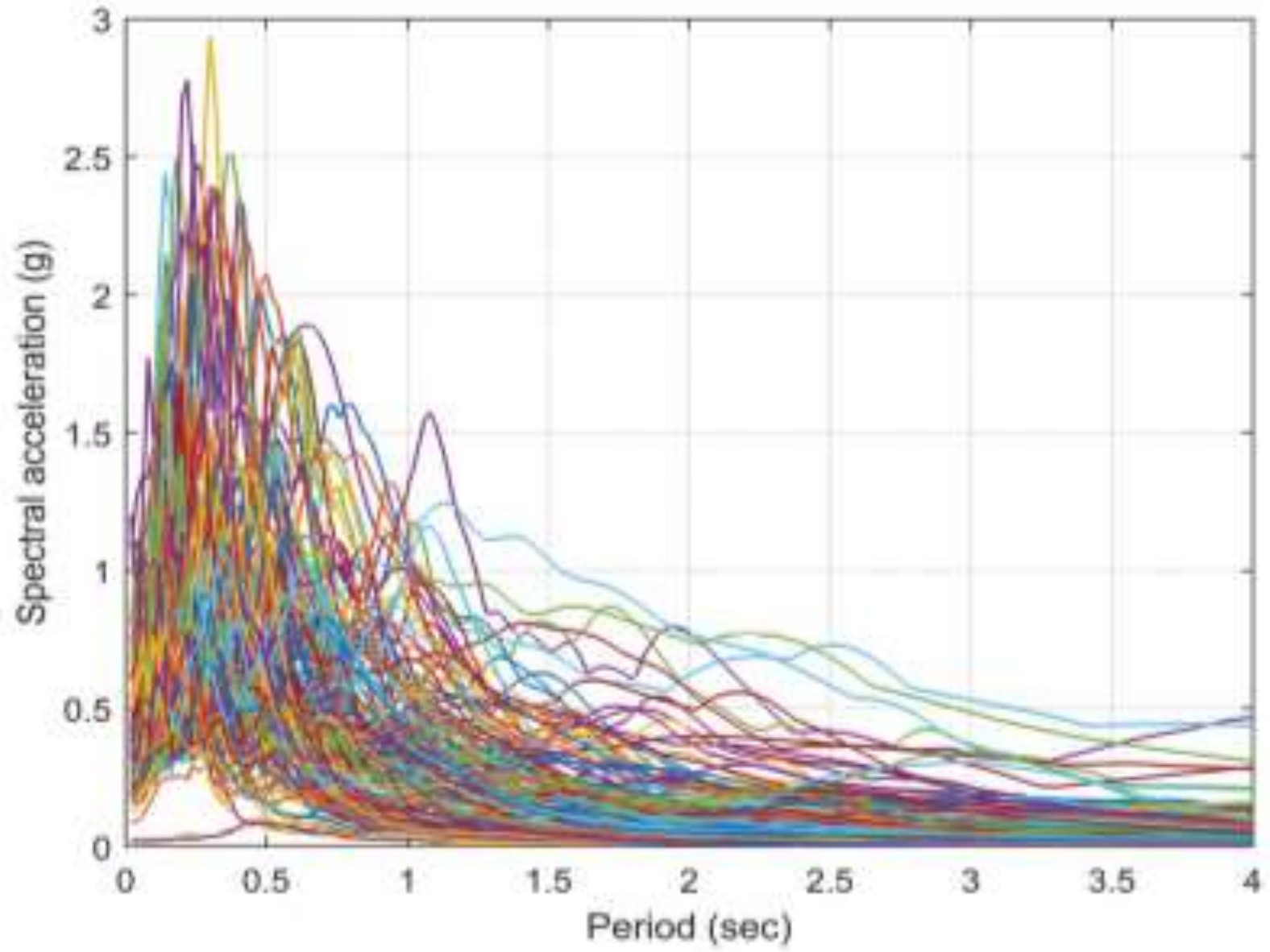

Figure 4. Acceleration spectra of 200 ground motions that are selected for a source of record selection procedure of this study

### 4.2 Spectral matching procedure

To remove and diminish the influence of spectral amplitudes of ground motions from the characteristics of the selected motions, earthquakes are matched to a target response spectrum. Hence, all of these motions only differ in terms of their duration as well as the non-stationary characteristics they inherited from original earthquake records. Spectral matching procedure modifies the original acceleration time history to match the entire range of target spectrum with minimal alteration of the velocity displacement history of the record. The time-domain spectral matching procedure proposed by Hancock et al. [32] is adopted in this study. The main assumption of this method is that the peak response does not change due to wavelet adjustment. Given N target spectral points to match, the spectral misfit is defined by the difference between the target spectral value ( $Q_i$ ) and the initial time series spectral value ( $R_i$ ):

$$\Delta R_i = \left(Q_i - R_i\right) P_i \tag{4}$$

where $P_i$ is the polarity of the peak response of the oscillator. Hancock et al. [32] shows that the response of an adjustment time series should be equal to $\Delta R_i$:

$$\Delta R_i = \sum_{j=1}^{N} b_j f_j(t) \qquad (5)$$

where $f_j(t)$ is a set of adjustment functions and $b_j$ is the set of amplitudes of the adjustment functions. The modified amplitude of the responses to the wavelet is determined not only by the misfit at each spectral point, but also by considering computed misfits at the neighboring spectral points:

$$b = C^{-1}\delta R \qquad (6)$$

where each component of a square matrix C is the amplitude of the wavelet response for the j-th spectral point at the peak oscillator time ($t_i$) of the initial time series response for i-th spectral point.

Acceleration and displacement median spectra can be extracted from the original time series and are depicted in Figure 5. As can be seen in this figure, acceleration and displacement spectra of the original (unmatched) ground motions have considerable dispersions and fluctuations. As far as the effect of motion duration is concerned, this source of variability should be minimized. In order to reduce this type of variability, the software SeismoMatch [33] is employed to match acceleration and displacement spectra of the selected ground motions to the predefined target spectra. As mentioned earlier, the method developed by Hancock et al. [32]—which has been already implemented in the above-mentioned software—has been used for the spectral matching procedure of this study.

In this study, the target spectra are the median of the acceleration and displacement spectra of the selected ground motions. Contrary to a design target spectrum which has sharp corners typically, the extracted target spectra (shown in Figure 5) are well-smoothed curves whose characteristics can improve the converging status of the spectral matching procedure. Besides, design response spectra are normally of Uniform Hazard Spectrum (UHS) type. Such response spectra are obtained in a way that they have the same probability of exceedance of spectral acceleration (SA) at all considered vibration periods, which are computed at a site considering all possible future earthquake events as visualized by the previous hazard scenario. On the contrary, a single earthquake event is quite unlikely to have such a response spectrum—i.e., of a UHS type.

For these reasons, we have decided to use median acceleration and displacement spectra for the spectral matching procedure. It is of the essence to say that consideration of the median response spectra as targets do not make a limitation for the proposed methodology framework. Because the spectrum-matched ground motions can be linearly scaled in such that their acceleration spectra at the structure's first mode period become equal to the code-based spectra which are obtained using different RPs at a relevant natural period.

The comparison regarding acceleration spectra of the original and matched time histories together with the associated target spectra is also presented in Figure 6. This figure indicates that the ground motions are well matched to the target spectra with a minimal change seen in the initial time histories of the ground motions. Acceleration spectra of matched ground motions versus the target spectrum of original ground motions, as shown in Figure 6 (a), demonstrate an acceptable match.

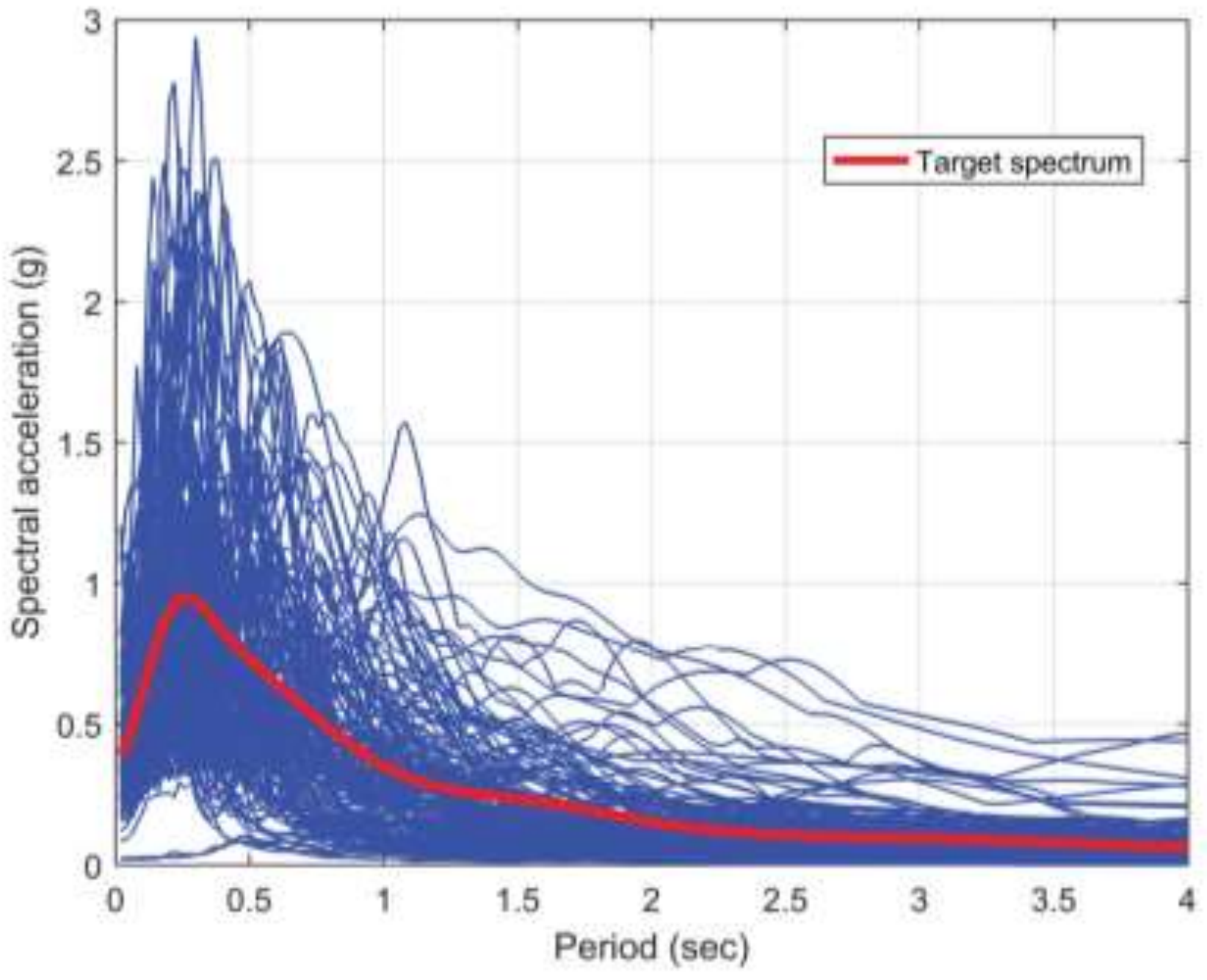

(a)

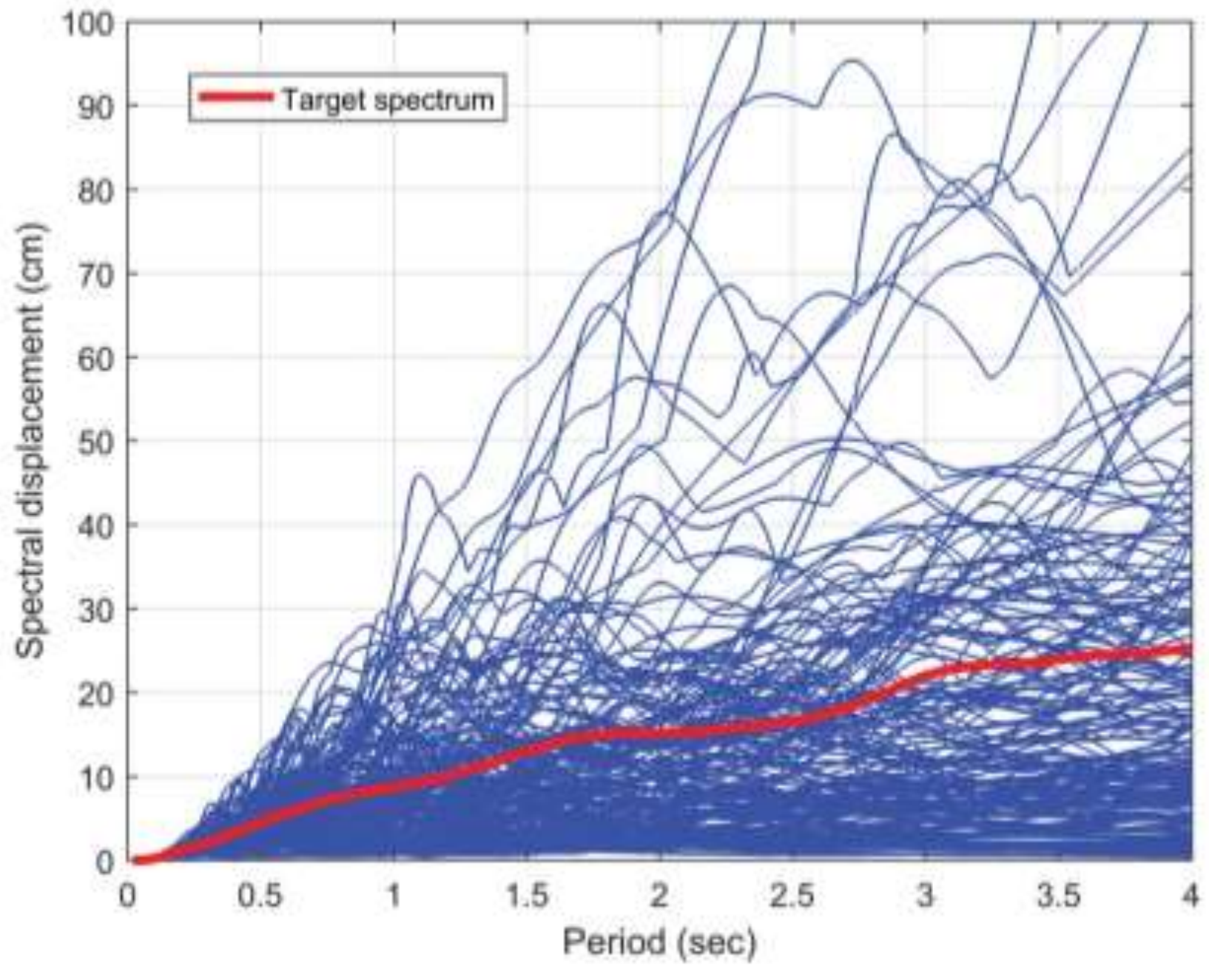

(b)

Figure 5. Spectral-related characteristics of the selected ground motions: a) acceleration spectra versus the target spectrum; b) displacement spectra and displacement target spectrum

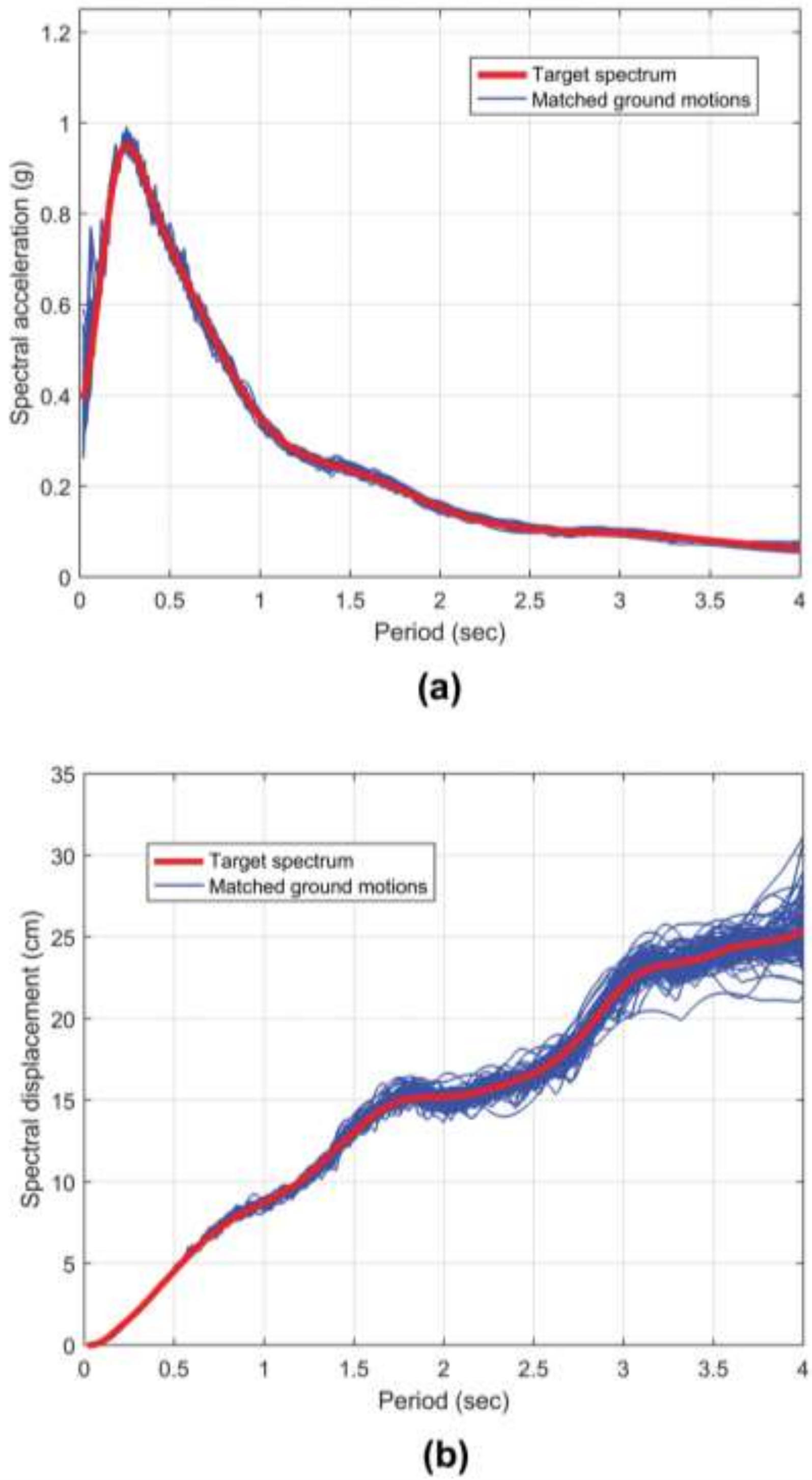

Figure 6. Spectral-related characteristics of adjusted ground motions: a) matched acceleration spectra versus the target spectrum; b) matched displacement spectra and displacement target spectrum

### 4.3 Dividing ground motions into long- and short- sets

In order to investigate the possible effects of motion duration on the seismic demands of the structures, spectrally equivalent ground motions are divided into two different sets based on the regulations of INBC in this regard. These specifications force structural and earthquake engineers to select earthquake records with strong motion duration equal to 10 seconds or more than 3 times of the fundamental period of the structure. In this case, the first set of motions includes earthquake records with significant strong motion duration between 10 to 15 seconds. And the second set of ground motions contains earthquake records with a significant duration equal or more than 15 seconds. The first group of motions can represent the short-duration earthquakes while the second one is related to a dataset that includes long-duration ground motions.

It is worth to add that there are different thresholds for the significant duration in the literature to pick short-duration ground motions in the record selection procedures. These thresholds can include: 5 seconds [34], 10 seconds [15], 20 seconds [4], 25 seconds [3,12,35] and 30 [36] as well as 35 [2] seconds. However, the geometrical mean significant duration of short-duration earthquakes in these studies is less than 15 seconds and varies from 6 (with a threshold of 20 seconds [4]) to 8.68 (with a threshold of 25 seconds [35]) and 14 seconds (with a 35-sec threshold [37]). In this study, the geometrical mean significant duration of the selected short-duration records is set to be 13.8 and is kept as close as possible to the corresponding threshold in INBC—i.e., the 10 seconds. Moreover, it is noteworthy to add that all of these ground motion records, both from short and long sets, are matched to a target response spectrum. Consequently, except for the duration-related sources of variability, these motions are almost unified in terms of amplitude-based intensity measures such spectral accelerations in different ranges of vibrational periods.

## 5 Structural modeling

The equivalent SDOF systems used for this study are created and modeled based on the bilinear pushover curves derived by Mashayekhi et al. [31], where four RC building type structures—three-, five-, eight- and twelve-stories—are numerically modeled and considered. In their study, the general characteristics of these structures including the number of stories, bay width, height length, and the total height of the structures are adopted according to the structural details reported by Korkmaz and Aktaş [38]. After that, the standard nonlinear static pushover curves of all considered RC structures are first computed and then converted to bilinear pushover curves. These curves can help to find the essential characteristics of the equivalent SDOF systems [24]. Figure 7 depicts standard as well as bilinear pushover curves for a structure with the model ID of 1008.

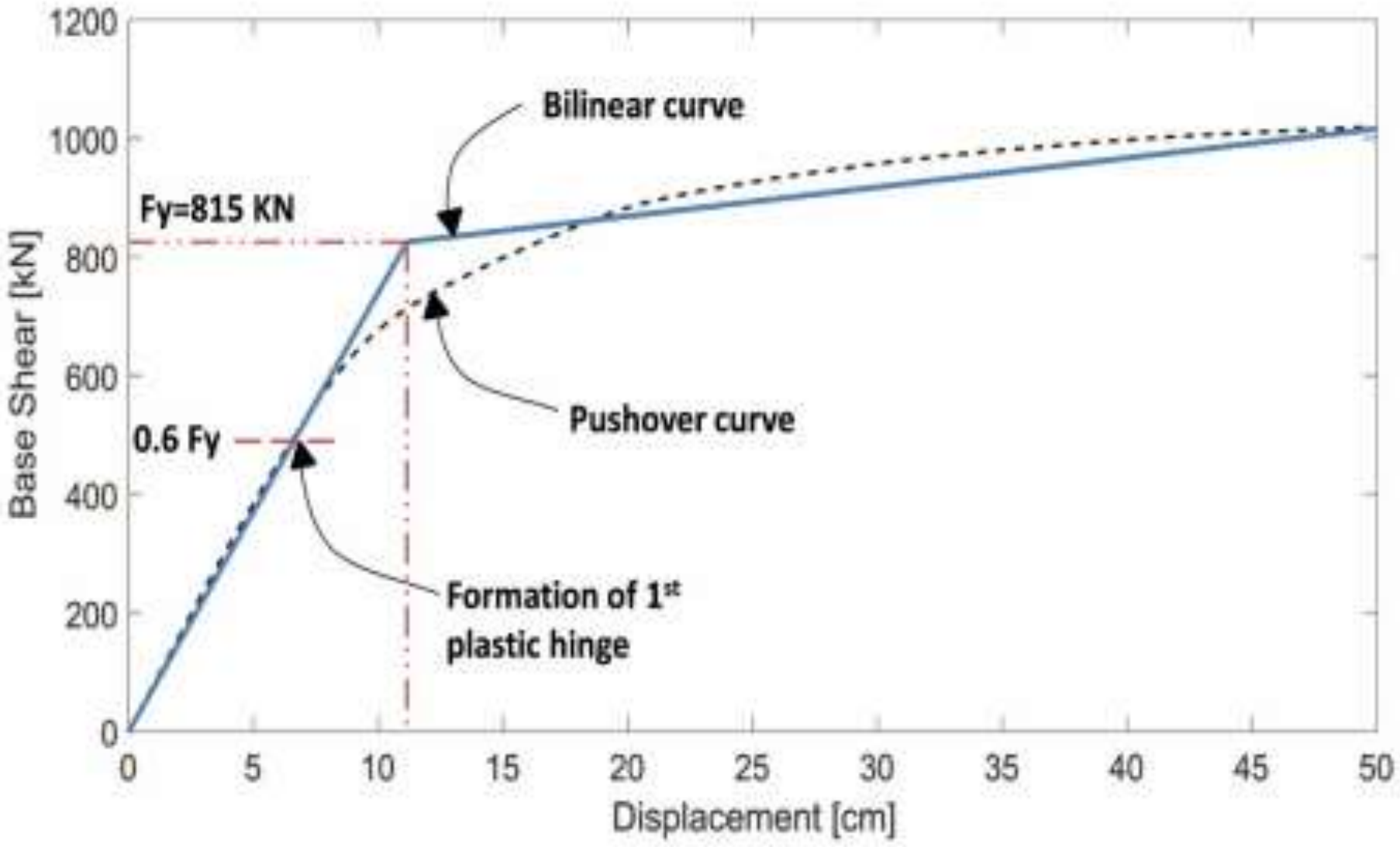

Figure 7. A standard pushover curve for a structure with model ID of 1008 and its bilinear equivalent to model the relevant equivalent SDOF [31]

Two types of equivalent SDOF system, one with degrading behavior and the other one without degradation, are modeled in Opensees software [39] to investigate the potential effects of motion duration on the seismic demands of the selected structures. In fact, for calculation of time history responses due to a dynamic input, we can consider the differential equation of motion for an SDOF system under earthquake excitations. The equation of motion in the incremental form is expressed as,

$$m\Delta\ddot{x}_i + c_t\Delta\dot{x}_i + \mathrm{k}_t\Delta x_i = -m\Delta\ddot{x}_g^i \qquad (7)$$

where $m$ is the mass of the SDOF system, $c_t$ is the tangent damping coefficient and $k_t$ is the tangent stiffness of the system, respectively. The term $\Delta\ddot{x}_g^i$ is also the increment of ground excitation time history.

The solution of the equation of motion for the SDOF system is obtained using the numerical integration technique such as the one suggested by Newmark [40]. The incremental quantities in Equation (7) are the change in the responses from time $t_i$ to $t_i + 1$. However, both the parameters related to the mass of the SDOF system ($m$) and the one related to the damping coefficient ($c_t$) are kept constant through the time history analysis. In this case, the equivalent mass of the SDOF systems ($m_e$) is calculated as the effective mass coefficient ($c_m$) in the first mode times the total mass of the N-story structures [24]. At the beginning of the time step, we should also have the initial values of the tangent stiffness and damping coefficient, the so-called $k_e$ and $c_e$:

$$K_e = \frac{4\,\pi^2 m_e}{T^2} \tag{8}$$

$$C_e = 2 m_e \omega \zeta \tag{9}$$

where T is the natural period of vibration of the equivalent MDOF, $\omega$ is the natural circular frequency which corresponds to the natural period of vibration with 2π/T and ζ is the viscous damping ratio of the system which is assumed to be 5% [9]. Table 1 shows the characteristics of the equivalent SDOF systems, which have been extracted from bilinear pushover curves of the considered MDOF structures.

Table 1. Characteristics of bilinear pushover curves required to model the equivalent SDOF structures

| Model ID | Fy (KN) | Seismic weight of mode 1, W ( MN) | Fy/W | Fundamental Period, T1 (sec) |
|---|---|---|---|---|
| 1003 | 500 | 2.04 | 0.24 | 0.61 |
| 1005 | 800 | 3.4 | 0.23 | 0.69 |
| 1008 | 815 | 5.24 | 0.1592 | 1.22 |
| 1012 | 1100 | 8.16 | 0.1374 | 1.4 |

In this study, structural SDOF systems with different periods of vibration—as specified in Table 1—are selected for structural modeling. The inelastic SDOF systems with degrading and non-degrading behavior are modeled using bilinear and Ibarra-Krawinkler hysteretic model [41], respectively. The employed hysteretic models for both degrading and non-degrading SDOF systems are displayed in Figure 8. These hysteretic models serve as a function for variation of tangent stiffness throughout the dynamic response history analysis. For degrading and non-degrading models, different $F_y/W$ parameters are applied as indicated in Table 1, where $W$ is the seismic weight of the equivalent SDOF systems and $F_y$ stands for the yield strength. Factors associated with the hardening as well as post-capping stiffness, $\alpha_s$ and $\alpha_c$, are selected to be 0.006 and -0.02, respectively. Residual strength is also 0.01 of the yield strength (λ=0.01).

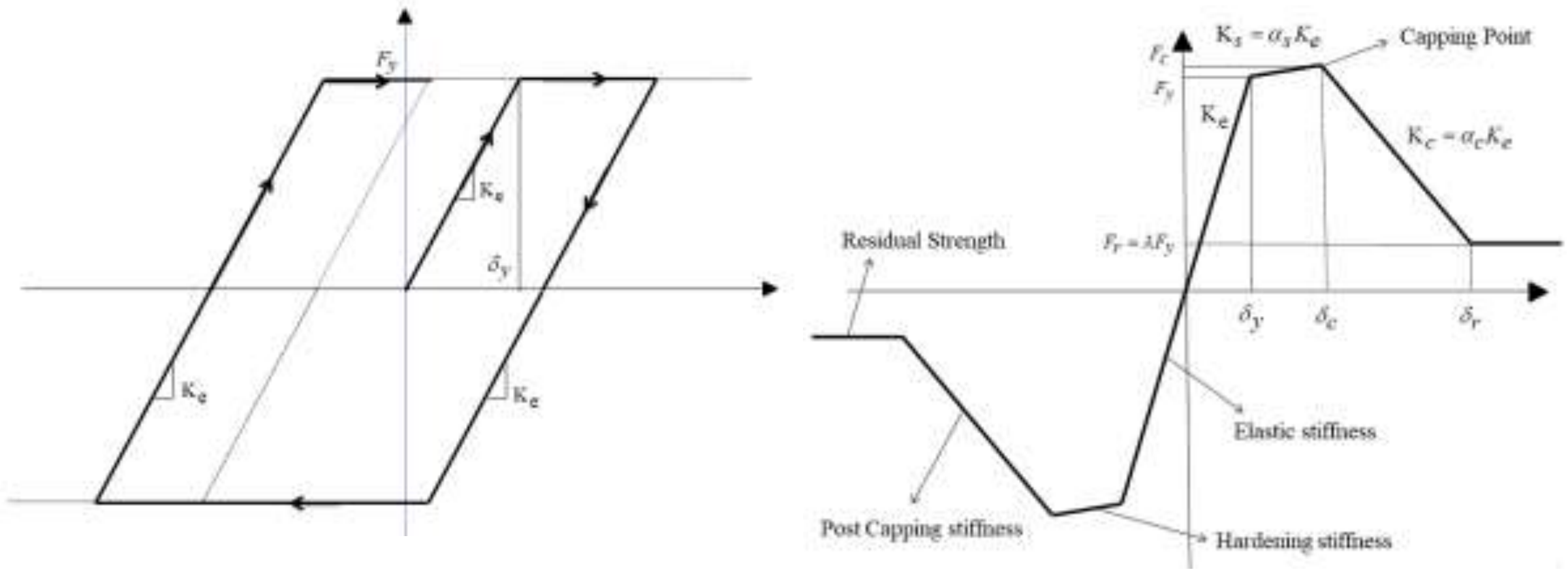

(a) (b)

Figure 8. Employed hysteretic models: a) bilinear elastoplastic model utilized for non-degrading SDOF system; b) backbone curve of the Ibarra-Krawinkler used for degrading SDOF system [41]

## 6 Numeral results

In this section, numerical results are presented. The response time history analysis essential for the statistical correlation procedure is accomplished at three seismic levels. The required seismic levels to study the correlation of motion duration with structural seismic demands at different levels of excitation are determined through earthquake RPs, as described in the research methodology section. For each structural system considered, the spectrum-matched ground motions are linearly scaled so that their acceleration spectra at the structure's first mode period become equal to the code-based spectra which are obtained using different RPs at the considered natural period. Next, the dynamic analysis of the structures subjected to the spectrum-matched ground motions, both from short- and long-duration sets, at multiple levels of excitation are performed through an IDA analysis. The results associated with the IDA procedure of all considered equivalent SDOFs are presented and compared to each other. It is worth mentioning that structural displacement demand is taken as the engineering demand parameter for both types of employed analyses, the IDA and time history analysis.

In order to quantify and get an insight into the relationship of motion duration and structural seismic demands, a statistical correlation procedure as described in the methodology section of this paper is performed at three levels of seismic excitation. As mentioned earlier, each of these seismic levels corresponds with an earthquake RP which is defined before. The 'Seismic Level 1' is chosen in a way that an RP of 100 years is considered, a seismic hazard level at SLE. The next seismic levels, the 'Seismic Level 2' and 'Seismic Level 3', are at the earthquake RPs of 475 (DBE) and 2475 (MCE) years, respectively. Correlation of significant duration with structural displacement for structure with the model ID of 1008, which is produced with a non-degrading SDOF model, are computed at three aforementioned seismic levels and shown in Figure 9 (a) to (c). The vertical red line in this figure demonstrates the threshold posed between short- and long-duration sets. As can be apparently recognized, the correlation of motion duration with structural displacement demand can increase at the upper seismic levels. For example, the correlation coefficient of significant duration and computed seismic structural demands is nearly equal to zero for the first considered seismic level, the Seismic Level 1.

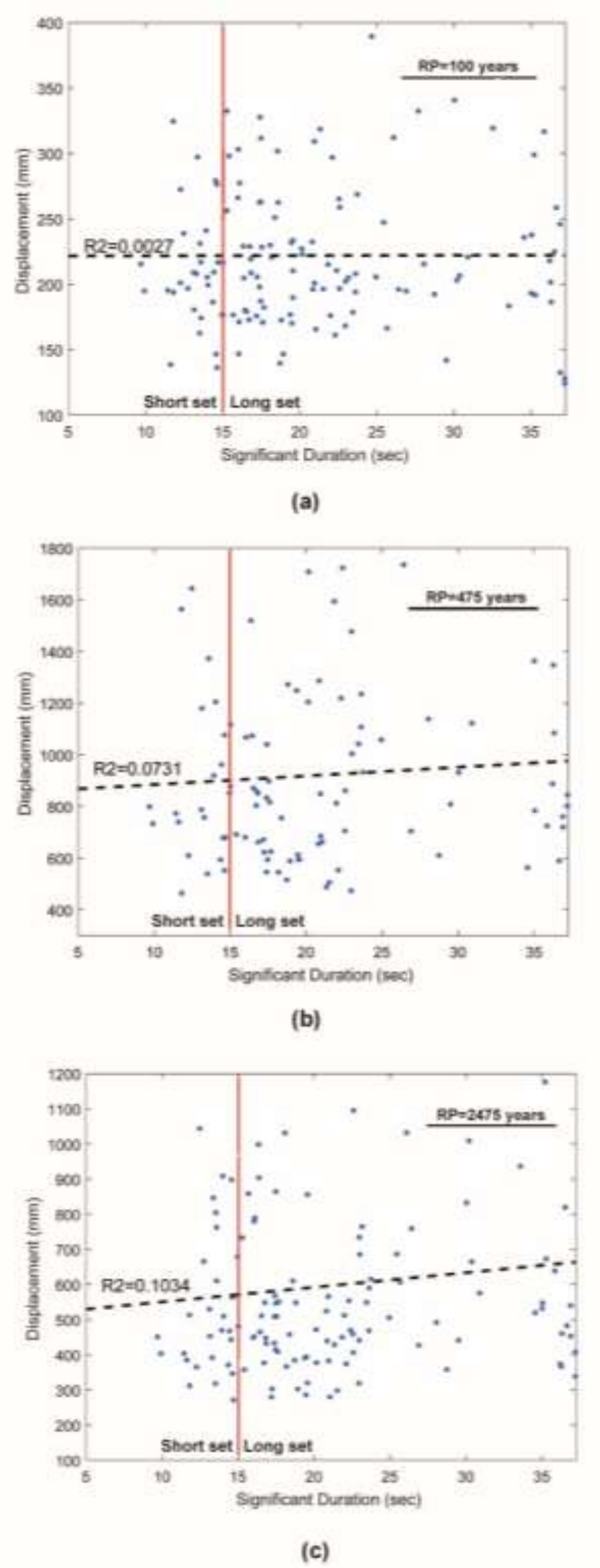

Figure 9. Correlation of motion duration with structural displacement for an 8-story non-degrading SDOF at three seismic levels: a) RP=100 years; b) RP=475 years; c) RP=2475 years

Correlation of motion duration and structural displacement of an 8-story structure (with a model ID of 1008)—which is represented by a degrading SDOF system—are depicted in Figure 10. A marginal increasing trend for correlation of motion duration and structural displacement demands, as observed in non-degrading SDOF model, can be seen in this structure with a degrading manner. In this case, the computed correlation coefficient goes up from 0.1103 to o.1632 from Seismic Level 1 (RP=100 years) to the Seismic Level 2 (RP=475 years), respectively. The correlation coefficients calculated for three considered seismic levels of this degrading model is more than the ones obtained for structures modeled without degradation. Therefore, the effect of motion duration on the displacement structural responses of degrading structures is more pronounced. However, it should be mentioned that the correlation coefficients obtained at multiple seismic intensity levels, the ones both from correlation studies of degrading and non-degrading structural systems, can confirm this matter that motion duration and structural displacement have a low positive correlation.

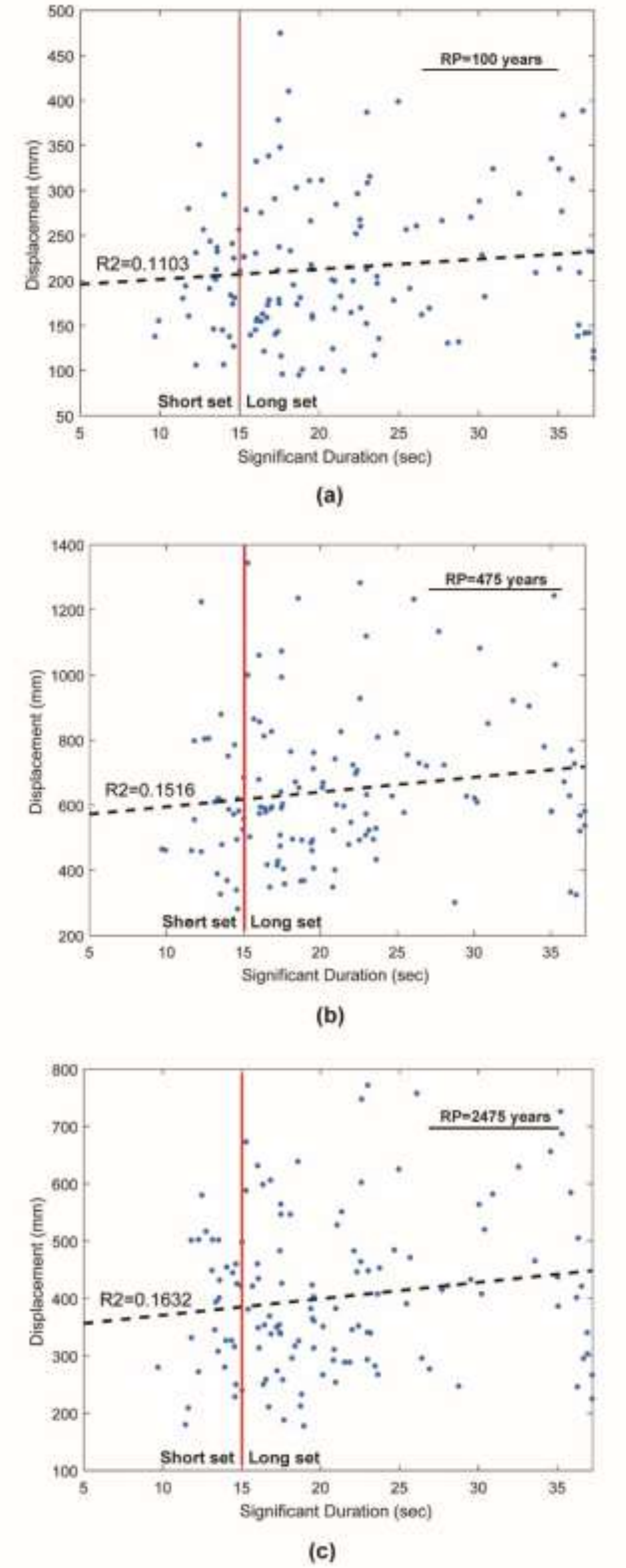

Figure 10. Correlation of duration with structural displacement for an 8-story degrading SDOF at three seismic levels: a) RP=100 years; b) RP=475 years; c) RP=2475 years

The same results, as described before for 8-story equivalent SDOFs, are found for the rest building types that are modeled and incorporated in this study. Correlation coefficients between duration and structural displacements of these structures are condensed and reported in Table 2 and 3. As can be seen in these tables, the same trend for correlation coefficients—as observed so far between duration and structural seismic demands of an 8-story building—is found for the rest buildings type models in use, the structures with 3 to 12 stories. In general, the correlation coefficients increase with an increase in the seismic level of the taken earthquake inputs. As indicated in Table 2, it is interesting to find out that the correlation coefficient in the Seismic level 1 (at the SLE) is negative for the non-degrading SDOFs with 3 and 12 stories.

Table 2. Correlation coefficients between duration and structural displacement demands of non-degrading (equivalent) SDOFs

| Seismic Level | 3-Story SDOF | 5-Story SDOF | 12-Story SDOF |
|---|---|---|---|
| 1 | -0.0915 | 0.00351 | -0.0151 |
| 2 | 0.0952 | 0.0951 | 0.1126 |
| 3 | 0.102 | 0.1249 | 0.1329 |

Table 3. Correlation coefficients between duration and structural displacement demands of degrading (equivalent) SDOFs

| Seismic Level | 3-Story SDOF | 5-Story SDOF | 12-Story SDOF |
|---|---|---|---|
| 1 | 0.0903 | 0.1042 | 0.1103 |
| 2 | 0.1311 | 0.1439 | 0.1429 |
| 3 | 0.1502 | 0.1527 | 0.1627 |

Figure 11 demonstrates single IDA curves, calculated for each applied ground motion, as well as the median response IDA curves of an 8-story frame which is modeled utilizing an equivalent non-degrading SDOF system. These response curves, shown in Figure 11 (a) and (b), are the outcomes of the IDA analyses that are performed up to a seismic level equal to the SA of 3 g for both of these equivalent SDOF systems, the degrading and non-degrading ones. Figure 12 also displays single IDA curves, which are computed for each applied ground motion, as well as the median response IDA curves of an 8-story RC structure with deteriorative manner. The single IDA curves for both sets of motions, the short and long sets of ground motions, are computed using an equivalent degrading SDOF system. The median response curves, obtained for each set of motions, represent the behavior of the considered structure under each group of applied ground shakings.

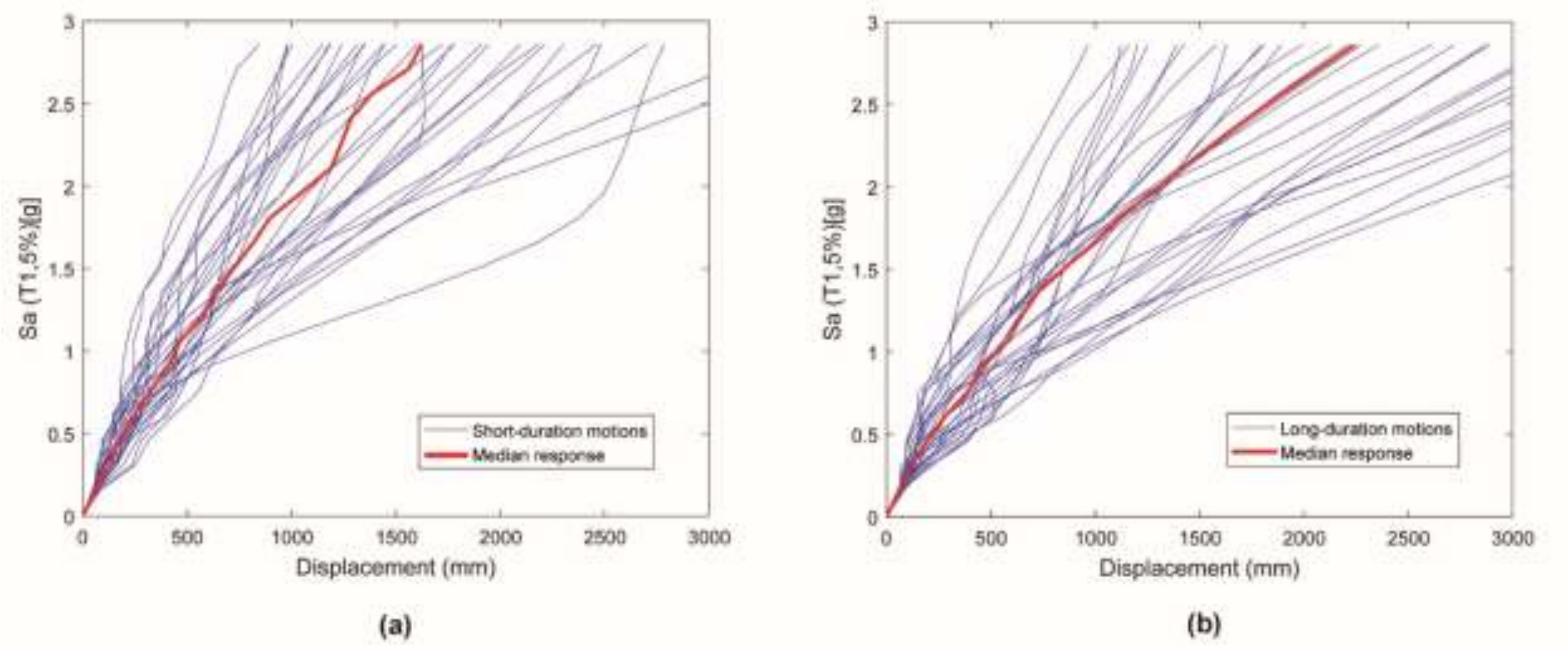

Figure 11. Single IDA curves and the median response of an 8-story structure using a non-degrading SDOF system: a) for short-duration motions b) for long-duration motions

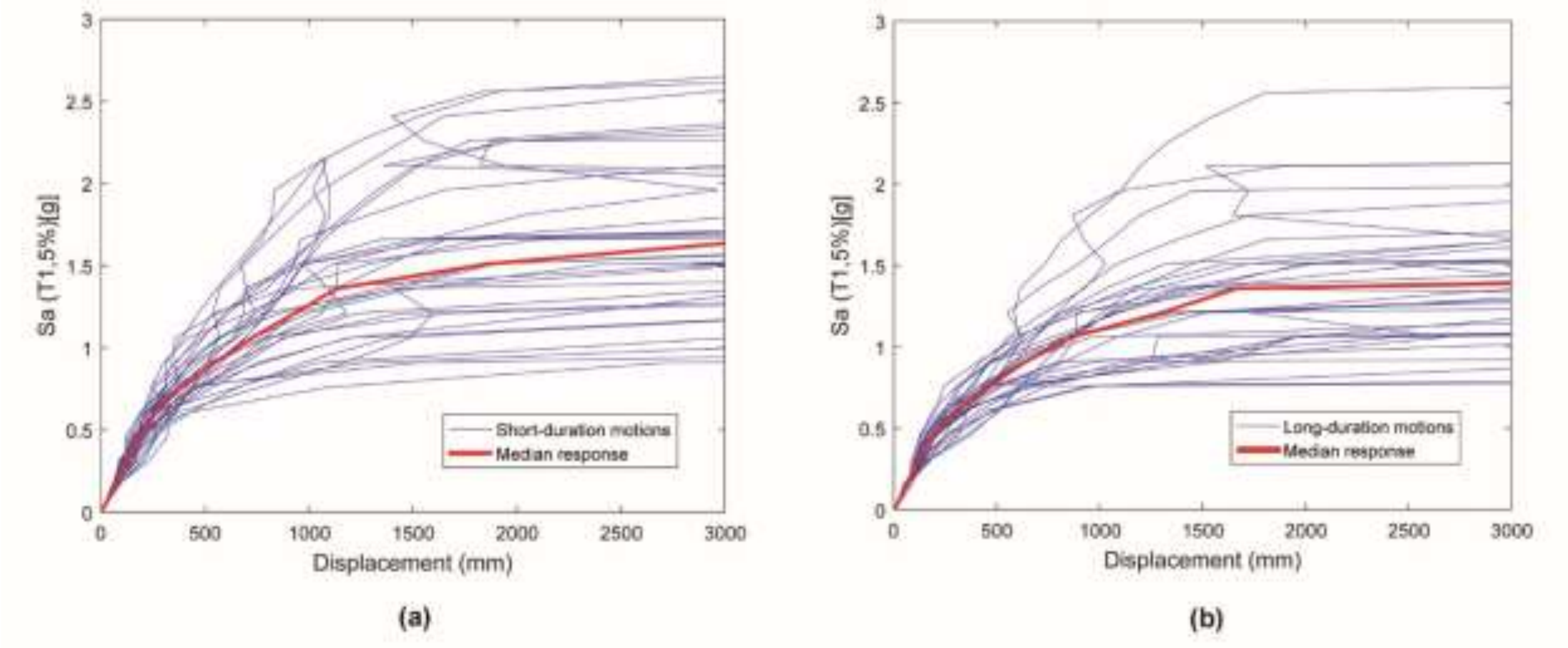

Figure 12. Single IDA curves and the median response of an 8-stort structure using a degrading SDOF system: a) for short-duration motions b) for long-duration motions

The median IDA response curves, for both sets of taken ground motions, are figured out and presented in Figure 13 for all employed equivalent SDOF structures. These SDOFs are created with the introduced non-degrading model and can be an appropriate response estimator for structures that do not have deteriorative manner. As can be witnessed in Figure 13 (a) to (c), at the lower levels of seismic excitation, the median IDA response curves for both sets of motions are the same while they get separated at the upper seismic levels once structures enter the regions associated with high nonlinear cycles. Contrary to the results obtained for structures under short-duration motions, structural seismic demands of non-degrading structures generally increase when they are subjected to long-duration motions. The increased seismic displacement observed in these structural systems, which are exposed to the long-duration set of ground motions, can be related to the further number of nonlinear cycles these building type structures experience during the long-duration earthquakes.

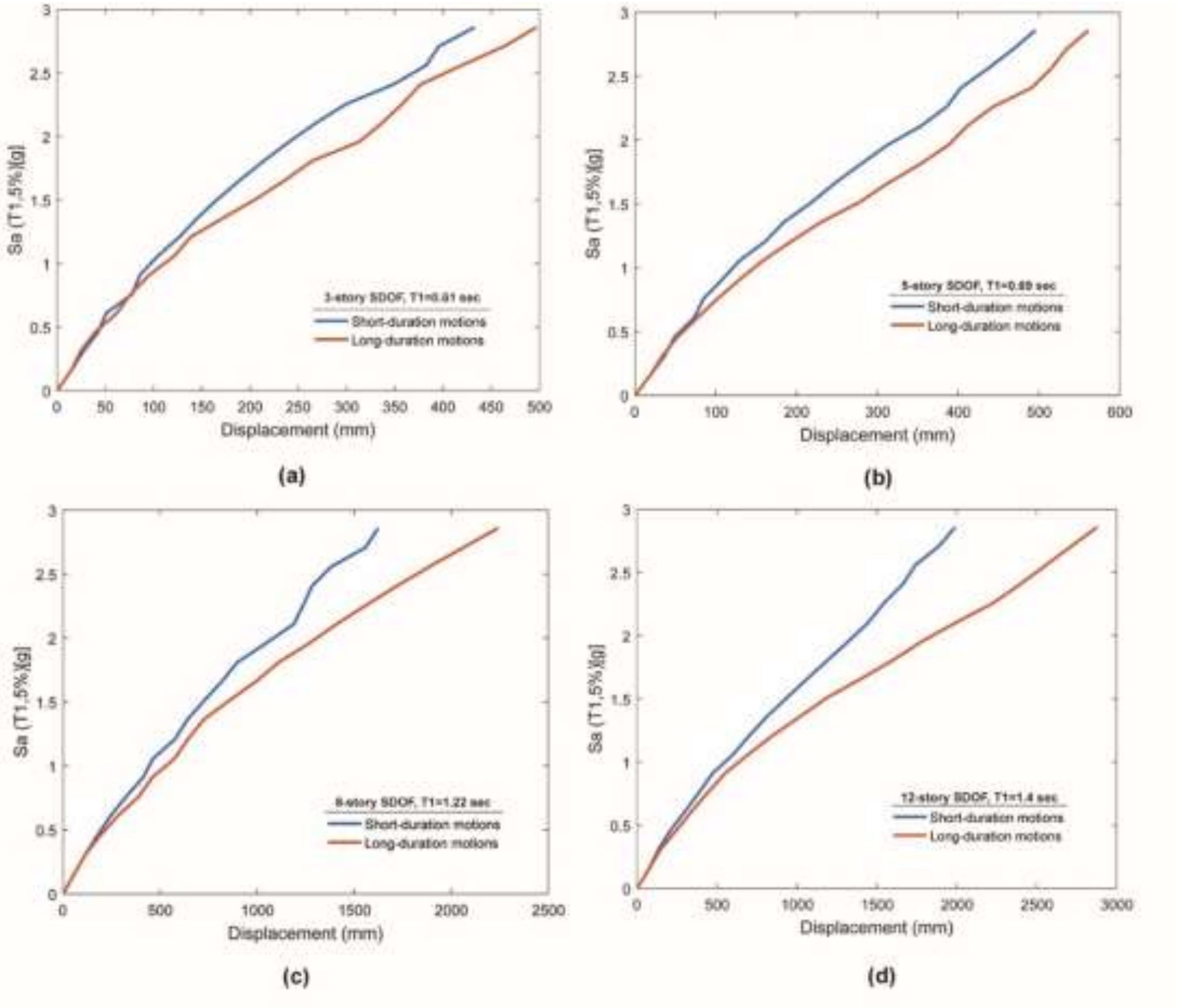

Figure 13. The IDA curves of short-duration motions versus the long-duration motions utilizing non-degrading SDOFs: a) for a 3-story building b) for a 5-story building c) for an 8-story building d) for a 12-story building

The median IDA response curves, for both sets of ground motions, are also computed and shown in Figure 14 for all considered equivalent degrading SDOF systems. As can be seen in Figure 14 (a) to (c), at the lower levels of seismic excitation, the median IDA response curves for both sets of ground motions are coincident while they detach from each other at the upper seismic levels. In these seismic levels, structures are under conspicuous nonlinear deformations. Therefore, given the fact that linear models are not capable of capturing the structural behavior in the nonlinear regions, they are not able to demonstrate the potential effects of motion duration on structural seismic demands. As a general rule, seismic demand imposed on the considered structural systems increases when they are exposed to long-duration motions, especially at the higher levels of seismic excitation. Also, the median IDA curves—computed for these two sets of motions—demonstrate a difference in the collapse levels obtained for each structural system. A collapse level in the IDA method is where a median IDA curve gets flatted or approaches a horizontal line at the upper seismic intensity levels. The median IDA curve obtained for the long-duration set of motions displays a clear reduction in the collapse level compared to the IDA curve calculated for the short set of motions. This reduction can reach 20 percent in the 5-story building type of this study as depicted in Figure 14 (b). It means that the spectral acceleration demand of this structural system, the 5-story building, show a 20 percent decline if it is subjected to long-duration motions. The observed behavior of these structural systems, exposed to long-duration motions, can be attributed to the more nonlinear cycles these structures typically experience compared to the buildings that may be subjected to a short set of ground motions. Moreover, the induced nonlinear cycles can weaken the structural members and thus further increase the associated peak deformation demands of the considered structures.

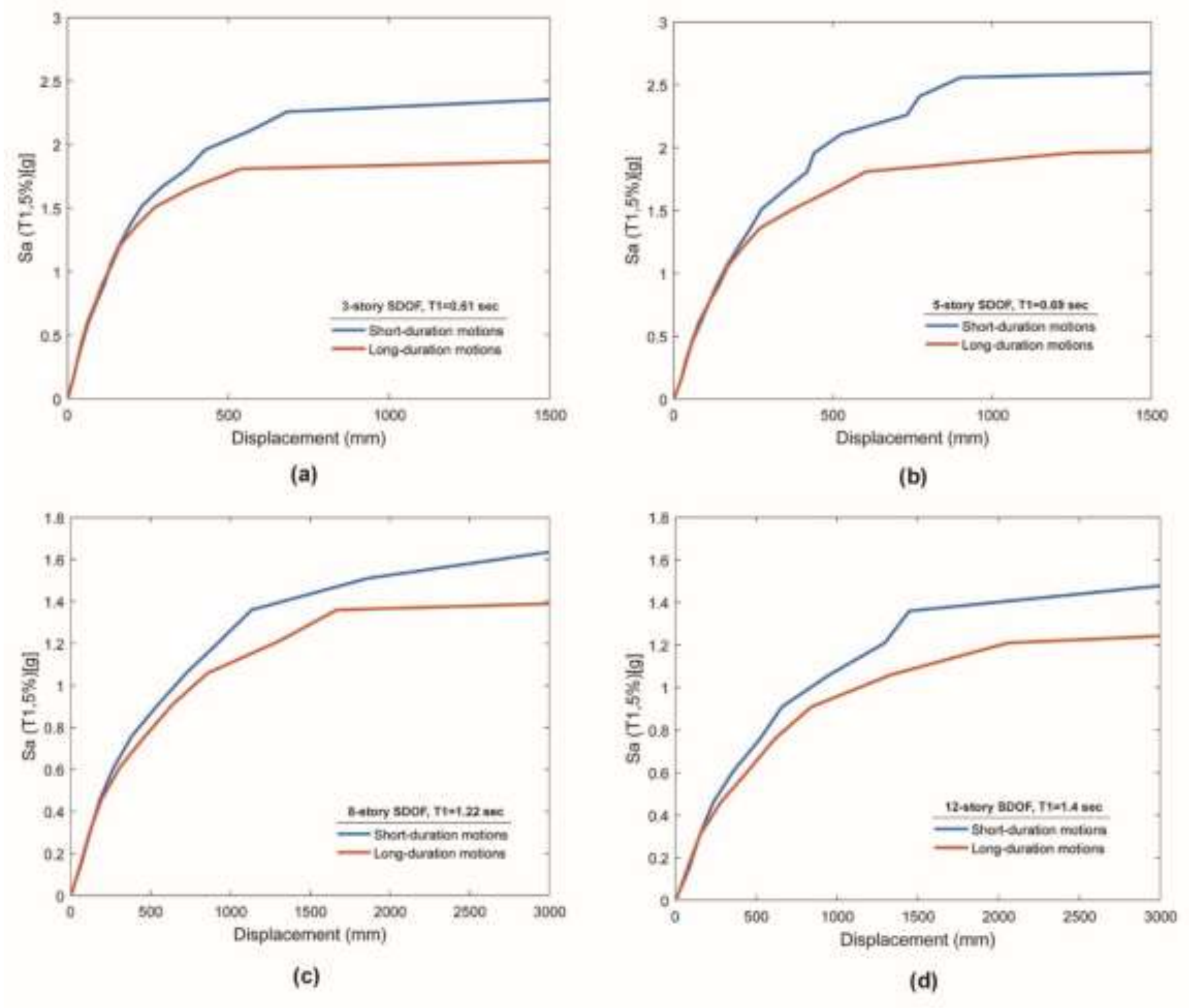

Figure 14. The IDA curves of short-duration motions versus the long-duration motions utilizing degrading SDOFs: a) for a 3-story building b) for a 5-story building c) for an 8-story building d) for a 12-story building

## 7 Discussion

Although we have found that duration and displacement may have a low positive correlation, it is demonstrated that ignoring the importance of the motion duration at the upper seismic intensity levels can have a considerable impact on the structural displacement demands of the buildings. We have shown that the collapse level in the IDA procedure gets changed if the category of the selected motions is altered from short- to long-duration. In this case, the collapse limit or dynamic instability is reached at a lower level of spectral acceleration demands once considered structures are subjected to the long-duration motions. It means that structural displacement demands at the upper levels of seismic intensity—or the collapse limit—may be overestimated in case short-duration earthquakes are selected for a site where long-duration earthquakes are more expected to happen. Therefore, in cases where an evaluation of collapse resistance of specific structural systems is considered, the proposed framework of this study can be hired to find out whether a structural type is sensitive to the duration of earthquakes.

The proposed framework of this study can be also used in structural reliability analysis [42]. In dynamic reliability analysis, a number of "explanatory functions" that control the structural responses of the building are chosen as potential candidates and considered as regressors in building response models of this approach. Based on the findings of this research, motion duration represents a means through which more damages can be expected and introduced into such models once longer motion durations are expected to appear at a specific site. Since explanatory variables in building response models are directly related to the structural and non-structural damages, a duration-related metric can thus serve as a potential regressor or explanatory function in developing new duration-dependent response models for each individual building.

Since it is revealed that inclusion of duration length may have significant impacts on the structural responses at high seismic intensity levels in IDA framework and considering the fact that guidelines such as Performance-Based Earthquake Engineering (PBEE) demand an accurate estimation of seismic responses very often [42,43], it is deduced that ground motion duration should be also regarded as the main criteria for the record selection procedures of current dynamic analysis frameworks. It is worth to mention that current record selection procedures are mainly based on the amplitude-based intensity measures [17,45]—i.e., working with PGA or SA at a range of natural periods of vibration. However, the duration-related measures have also the advantage of being easily estimated for applications from predictive models (e.g. Afshari and Stewart [46]) developed for estimation of the record-based duration as a function of magnitude, distance, site classification and other explanatory variables. Therefore, for a more reliable estimation of seismic responses, one can then easily make use of such predictive models in seismic risk assessments.

# 8 Conclusion

This paper examines the influence of applying different seismic levels on the results of correlation-based assessments conducted between motion duration and structural seismic demands. Structural seismic demands are determined through an IDA as well as a nonlinear time history analysis. Spectrally matched ground motions are employed in these analyses to investigate the potential effects of motion duration on the structural responses, isolating the contribution of earthquake duration from the effects of ground motion amplitudes and response spectral shape. For computing linear correlation coefficients between motion duration and structural response, three seismic levels are determined in a way that each of them is compatible with an earthquake return period (RP). Four single degree of freedom systems, which are derived from four real reinforced concrete structures, are considered to model required equivalent SDOFs—degrading and non-degrading models. The results are listed below:

- Although it seems obvious to find out that correlation coefficients approach to zero or even negative value in non-degrading systems at low RPs, small positive correlation coefficients equal to 10 percent have been witnessed in degrading SDOFs at the same seismic levels.

- It is revealed that correlation of motion duration with structural seismic demands do not remain unchanged and increase with earthquake RPs where more nonlinearity is expected to happen in the selected structures, both for degrading and non-degrading SDOFs.

- It is revealed that long-duration motions can cause up to 50% larger peak deformation demands compared to the corresponding seismic demands imposed by the short-duration ground motions.

- It is shown that a reductive trend in collapse level of the median IDA curves of the degrading SDOFs is found when they are exposed to long-duration ground motions. At the collapse limit, the spectral acceleration demand of structures under such long excitations can get declined by 20 percent in some cases.

## Acknowledgment

The authors would like to thank three anonymous reviewers for their helpful and constructive comments that greatly contributed to improving the final version of the paper. We also thank all the efforts accomplished by the staffs in the center of High-Performance Computing (HPC) of the Sharif University of Technology for providing a robust and fast platform to run our simulations of this research.